\documentstyle[aps,draft]{revtex}

\begin{document}

\title{Acceleration, streamlines and potential flows in
 general relativity: analytical and numerical results.}

\author{Maximiliano Ujevic\thanks{e-mail: mujevic@ifi.unicamp.br}}
\address{Instituto de F\'{\i}sica ``Gleb Wataghin'' \\ Universidade Estadual de
Campinas, 13083-970, Campinas, SP, Brasil}
 \author{Patricio S. Letelier\thanks{e-mail: letelier@ime.unicamp.br}}
\address{Departamento de Matem\'atica Aplicada, Instituto de Matem\'atica,
Estat\'{\i}stica e Computa\c{c}\~ao Cient\'{\i}fica \\ Universidade
Estadual de Campinas,  13081-970, Campinas, SP, Brasil}


\maketitle

\begin{abstract}
Analytical and numerical solutions for the integral curves of the
velocity field (streamlines) of  a steady-state  flow of an ideal
fluid with $p = \rho$ equation of state are presented. The
streamlines associated with an  accelerate black hole and a rigid
sphere are studied in some detail, as well as, the
 velocity fields of  a black hole and a rigid sphere  in an external
 dipolar field (constant acceleration field). In the latter case the dipole
  field is produced by an axially symmetric halo or shell of matter. For
each case the fluid density is studied using contour lines. We found that
the presence of acceleration is detected by these contour lines.
 As far as we know this is the first time
 that the integral curves of the velocity field for
  accelerate objects and related
 spacetimes are studied in general relativity.

\end{abstract}


\section{Introduction}

The study of potential flows in general relativity
is  relevant in the understanding of several phenomena
 of interest in relativistic astrophysics like: fluid flows at
 relativistic speeds
  in the presence  of  a  neutron star \cite{sha,abr:sha}, star
  clusters moving in gaseous
  media \cite{sha},
  flows near a cosmic string \cite{sha}, accretion in binary star systems
  and supermassive black holes \cite{pet:sha:teu}, and
   others \cite{sha:teu,fei:iba,fon:mar:iba:mir}. Also, numerical
solutions of the equations of general relativistic hydrodynamics can
simulate, and model, gravitational collapse and the evolution of neutron
stars \cite{fon}.

   Most of the  articles in this area deal with
    the important
case of  fluid motion evolving in the spacetime associated with
 compact stars and black holes. The
implementation of new background metrics
brings some new challenges.  First, metrics other
than Schwarzschild and Kerr
 are not so well studied, sometimes  a complete understanding
 of the physical
  meaning of the metric  is missing.  Also,  the
  solutions of
  the fluid equations
  in a non trivial metric may be quite involved. In particular,
  the search for
   significant boundary values (or initial conditions) presents
   a non trivial
  problem. A simple and paradigmatic
case of a flow is the stationary-zero-vorticity flow   of a fluid
 with  adiabatic
stiff equation of state. In this case, for  relativistic flows,
the fluid equations admit  analytical
 solutions for some particular metrics \cite{sha,pet:sha:teu}. These solutions
are  used as  test-beds for testing almost all the numerical
hydrodynamic codes in the subject. Other
 potential flows in a  non stationary background
and different equation of state have been studied, see for
instance \cite{abr:sha}. Also, the solutions for potential flows
permit to test new optimized codes in resolving nonlinear
hyperbolic systems of conservation laws \cite{fon:mar:iba:mir}, see for a
representative sample of numerical schemes \cite{fon}.

In this work we extend the investigations about potential flows by studying the
streamlines of a steady-state ideal fluid in the presence of an accelerated black hole
and rigid sphere, and of a black hole and rigid sphere with a dipole field produced by
an axially symmetric halo of matter. This shell like structures are useful in modeling
many situations of interest in astrophysics, as for example, the Supernova 1987A
\cite{pan:etal,mey,che}, for other applications see \cite{vie:let}. We assume that the
fluid is a test fluid, i.e., the metric does not evolve and it is given $apriori$. The
state equation and the idea of a rigid star (rigid sphere) we use are idealized.
However, they bring important results about the instability and behaviour of the
fluid, and the difficulties involved in this kind of scenarios.

  This work is divided as follows. In Sec. II we present the basic equations that describe
  potential flows. In Sec. III we summarize some aspects of
  the Weyl C-metric that
represents the spacetime associated with an uniformly accelerated
black hole. In particular,
 we present the metric in different  systems of
 coordinates to facilitate the
 physical interpretation of the results.
In the subsections  III.A and III.B we study
 potential flows for an accelerated black hole
and an  accelerated rigid sphere, respectively. In the
 first case we use a perturbative approach and in
 the second we solve numerically the potential equation. For
 both cases, we
also study the behaviour of the fluid density. In Sec. IV we
introduce a metric that represents a black hole in a dipolar
field, this field is produced by an external halo of matter. In
 the subsections  IV.A and  IV.B, respectively, we study
the streamlines of a fluid in the presence of a rigid sphere and black hole, both with
halo. Also, in both cases we study the fluid density. Finally in Sec. V we summarized
our results.

\section{Basic Equations}

The starting point of this work is an ideal fluid whose
energy-moment tensor is given by $T_{\mu\nu}=(\rho+p) U_{\mu}
U_{\nu} + p g_{\mu\nu}$, where $p$ is the pressure, $\rho$ the
total energy density and $U_\mu$ the four-velocity. The
conservation
 equations,
$T^{\mu\nu}_{;\nu} =0$, for this kind of fluid reduce to

\begin{equation}
(\rho + p) U^\mu_{;\mu} + \rho_{,\mu} U^\mu =0,
\end{equation}

\noindent and

\begin{equation}
(\rho +p) U^{\nu} U_{\mu ; \nu} + p_{,\mu} +
 p_{,\nu} U^\nu U_\mu= 0, \label{euler}
\end{equation}

\noindent which are respectively the conservation an Euler equation. (Our conventions
are: $G=c=1$. Metric with signature $+2$.
 Partial and covariant  derivatives with respect
to the coordinate $x^\mu$   denoted by  $,\mu$ and $;\mu$,
  respectively.)
For isentropic
 flows we have,
$(\sigma/n)_{;\mu}=0$, where $\sigma$ is the
 entropy per unit volume and $n$ the
baryon number density. In this case the equations of motion (\ref{euler}) take the
form \cite{lan:lif},

\begin{equation}
U^\nu \omega_{\mu\nu} = 0,
\end{equation}

\noindent where $\omega_{\mu\nu}$ is the relativistic
 vorticity tensor defined as

\begin{equation}
\omega_{\mu\nu}=\left[ \left( {{\rho+p}\over{n}} \right) U_\mu \right]_{;\nu} -
\left[ \left( {{\rho+p}\over{n}} \right) U_\nu \right]_{;\mu}.
\end{equation}

\noindent The potential flow  solution of this equation
 ($\omega_{\mu\nu}=0$) is

\begin{equation}
\left( {{\rho+p}\over{n}} \right) U_\mu = \Phi_{,\mu}, \label{potflow}
\end{equation}

\noindent  where $\Phi$ is a scalar field. From (\ref{potflow})
and  the equation of continuity for the baryon number density $n$,
$(n U^\mu)_{;\mu}=0$, we obtain the differential equation for
 the scalar field $\Phi$,

\begin{equation}
\left[ \left( {{n^2}\over{\rho+p}} \right) \Phi_{,\nu} g^{\nu\mu}
\right]_{;\mu}=0. \label{eqphi}
\end{equation}

\noindent The normalization condition, $U_\mu U^\mu=-1$, provides
a relation between the pressure, the total energy density, and
scalar field: $(\rho+p)/n = \sqrt{-\Phi_{,\mu} \Phi^{,\mu}}$.
Also,  $p$ and $\rho$ are assumed to be related by a barotropic
 equation of state, $p=p(\rho)$. In general, Eq. (\ref{eqphi}) is a
 nonlinear equation for the scalar field $\Phi$. However it becomes linear
if we assume that $p=\rho \propto n^2$, i.e., a stiff equation of
state. Thus,

\begin{equation}
n U_\mu = \Phi_{,\mu}, \;\;\; n^2= - \Phi_{,\sigma} \Phi^{,\sigma},
\label{velocity}
\end{equation}

\noindent  and $\Phi$ is a solution of

\begin{equation}
\Box \Phi \equiv (\sqrt{-g} g^{\mu\nu} \Phi_{,\mu})_{,\nu} = 0,
\label{diffeq}
\end{equation}

\noindent  which is the usual  wave equation for a scalar field.
 Note that for the stiff
 equation of state the sound
velocity in the fluid is equal to the velocity of light, therefore
the flow is always subsonic and the presence of shockwaves is
excluded.

\section{Accelerated black holes and rigid spheres}

The Weyl C-metric is the member of the static
axially symmetric Weyl
 family of
solution of the vacuum Einstein equations \cite{kin:wal},
\begin{equation}
ds^2={1\over{A^2(q+p)^2}} \left[- F(q) dt^2 +
{{dp^2}\over{G(p)}} +
{{dq^2}\over{F(q)}} + {G(p)}
d\varphi^2 \right], \label{cmetric}
\end{equation}
 where the functions $G(p)$ and $F(q)$ are
 the cubic polynomials,
\begin{eqnarray}
&&F(q) = -1 + q^2 -2 m A q^3, \nonumber \\
 &&G(p)= 1 - p^2 -2 m A p^3. \label{poly}
\end{eqnarray}
The coordinates $t$, $p$, $q$ and $\varphi$ are dimensionless, $A$
is a  constant with dimension of $L^{-1}$. The range of $t$ is
  $(-\infty,+\infty)$, while the
range of $\varphi$ is $[ 0 , 2\pi]$. The
ranges of the coordinates $p$ and $q$
depend on the roots of $F$ and $G$. The constraint
 $m^2 A^2 < 1/27$ is imposed
to ensure the existence of three real roots in (\ref{poly}). This
condition is not a physical one, it is only due to
the choice of coordinates \cite{let:oli}. From the fact that the functions
have three real roots, the C-metric allows a description
of different spacetimes \cite{let:oli} depending
on the interval considered for ($q,p$). In this work the
functions $F(q)$ and $G(p)$ are
bounded within the zeros $(q_R,q_S)$ and $(p_0,p_\pi)$
 given in \cite{far:zim}. In this
case , $m$ and $A$ are interpreted as the mass and
 acceleration of a black
hole \cite{bon}. A  Newtonian image of the  matter content
of this metric
 is given by a rod and a semi-infinite line mass,
  both of
density $1/2$ and placed along the $z$-axes. The metric
 is interpreted
 as the representation of particle (the rod) in a accelerating
 frame (the semi-infinite line), in other words,
associated to the spacetimes of Schwarzschild and Rindler. The
rod and the semi-infinite line has  a strut
holding them apart that represents  a conical singularity.
We note that the metric (\ref{cmetric}) diverges when the
acceleration is zero ($A\rightarrow 0$). Hence, in this coordinates, we
 do not
have the correct limit that corresponds to the Schwarzschild
metric. To  obtain this last metric when $A\rightarrow 0$ we  make
the coordinate transformation \cite{far:zim},
\begin{equation}
r = {1\over{A(q+p)}}, \;\;\; t \rightarrow At. \label{trans}
\end{equation}
Now the metric (\ref{cmetric}) reads,
\begin{equation}
ds^2 = - H dt^2 + {1\over{H}} dr^2 +
{{2 A r^2}\over{H}} dr dp + r^2 \left(
{1\over{F}} + {1\over{G}} \right) dp^2 + r^2 G d \varphi^2,
\label{modcmetric}
\end{equation}
 where
\begin{eqnarray}
H&=&A^2 r^2 F \nonumber \\
&= &-A^2 r^2 G(p - A^{-1} r^{-1}) \nonumber \\
&= &1 - {{2m}\over{r}} + 6 A m p + A rG_{,p} - A^2 r^2 G(p).
\end{eqnarray}

When $A \rightarrow 0$  the metric (\ref{modcmetric}) reduces to Schwarzschild form
provided that the spherical angular coordinate $\theta$ be related to $p$ by $G(p)
=1-p^2 = \sin^2 \theta$. Furthermore, if the mass is zero in (\ref{modcmetric}) ( $m
\rightarrow 0$) the space becomes Euclidean with a special form of a flat space in an
uniformly accelerated frame \cite{far:zim}. For these reasons the line element
(\ref{modcmetric}) represents a uniformly accelerating Schwarzschild-type particle.
The coordinates ($t,r,p,\varphi$) is a coordinate system rigidly fixed on the
accelerating particle. With the transformation (\ref{trans}) we gain the correct
Schwarzschild limit,  but now the metric is not diagonal.

In the case of $A\neq 0$, the angular coordinates $p$ and $\theta$ are
related  by  $G(p) =\sin^2 \theta$. The mapping between them is
\cite{far:zim},

\begin{equation}
p = \left\{
      - \frac{1}{6Am} \left[
       2 \cos \left( \frac{\Theta(\theta)}{3} +
       \frac{4\pi}{3} \right) + 1 \right] \; {\rm for}
       \; 0 \leq \theta \leq
      \frac{\pi}{2}, \\
        - \frac{1}{6Am} \left[ 2 \cos \left(
      \frac{\Theta(\theta)}{3} +
       \frac{2\pi}{3} \right) + 1 \right] \; {\rm for}
       \; \frac{\pi}{2} \leq \theta \leq \pi,
      \right.
\end{equation}
 where
\begin{equation}
\cos \Theta(\theta)= 1-54 A^2 m^2 \cos^2 \theta.
\end{equation}

Due to the acceleration the black hole
horizon deforms.  The exact form of
this horizon is \cite{far:zim},
\begin{equation}
r_{Sch} = \left\{
      - {{3m}\over{\cos \left( {{\Theta(\theta)}\over{3}} + {{4\pi}\over{3}} \right)
        + \cos \left( {\delta\over{3}} + {{2\pi}\over{3}} \right)}} \; {\rm for}
        \; 0
        \leq \theta \leq {\pi\over{2}},\\
 - {{3m}\over{\cos \left( {{\Theta(\theta)}\over{3}} + {{2\pi}\over{3}} \right)
        + \cos \left( {\delta\over{3}} + {{2\pi}\over{3}} \right)}} \; {\rm for}
        \; {\pi\over{2}} \leq \theta \leq \pi.
    \right.
\label{horizon}
\end{equation}
 The form of the Schwarzschild horizon will be used in a
perturbation scheme to place an upper limit on the acceleration.

\subsection{Potential flows and accelerated black holes}

In order to study streamlines of the flow we need to know the field
$\Phi$ that
 determines the fluid velocity.
 The explicit form of the  equation for $\Phi$ can
 be obtained from  (\ref{diffeq}) and  (\ref{modcmetric}). We find,
\begin{equation}
\Box \Phi = - {{r^2}\over{H}} \Phi_{,tt} +
( [H + A^2 r^2 G] r^2 \Phi_{,r} )_{,r}
+(G \Phi_{,p} )_{,p}  + {1\over{G}}
\Phi_{,\varphi\varphi} - (r^2 G A \Phi_{,r} )_{,p}
- (r^2 G A \Phi_{,p})_{,r} = 0. \label{boxcm}
\end{equation}

We shall assume that the fluid is stationary and also independent of the variable
$\varphi$. In this case the time dependence of $\Phi$ will be set by adding the term
$-at$ to the final solution, where $a$ is a constant, this term clearly satisfies
(\ref{boxcm}). The arbitrary constant $a$ is related to the $0^{th}$ component of the
four-velocity. With these assumptions, and setting $m=1$, (\ref{boxcm}) takes the form
\begin{equation}
\Box \Phi = \Box_{Sch} \Phi + A \Box_1 \Phi + A^2 \Box_2 \Phi =0,
\label{boxcm2}
\end{equation}
 with
\begin{eqnarray}
&&\Box_{Sch} \Phi = ([r^2-2 r] \Phi_{,r} )_{,r}
+ ([1-p^2] \Phi_{,p})_{,p}, \nonumber
\\ \nonumber
\\ &&\Box_1 \Phi = ([6 p r^2 - 2 p r^3] \Phi_{,r} )_{,r}
 - (2 p^3 \Phi_{,p})_{,p}
 - (r^2 [1-p^2] \Phi_{,r})_{,p} -
 (r^2 [1-p^2] \Phi_{,p})_{,r},
\\ \nonumber
\\ &&\Box_2 \Phi = -(6 r^3 p^2 \Phi_{,r})_{,r}
+ (2 r^2 p^3 \Phi_{,r} )_{,p} +
(2 r^2 p^3 \Phi_{,p} )_{,r}. \nonumber
\end{eqnarray}

Finding an analytical solution to Eq. (\ref{boxcm2}) does not appear
possible. We shall look for a  meaningful approximate solution.
 For small enough  $A$,  say $A \leq 0.01$, the Schwarzschild surface
remains almost unaltered [cf. (\ref{horizon})]. Then we can
approximate (\ref{boxcm2}) as
\begin{eqnarray}
\Box \Phi = \Box_{Sch} \Phi + A \Box_1 \Phi =0. \label{boxcm3}
\end{eqnarray}
By using the fact that $A \approx 0$ we may solve this equation
 by a perturbative method,
but first we  recall some earlier results. The solution for a
black hole when $A=0$ with the condition that at infinity the
fluid velocity is
 constant and parallel to the $z$-axes of
the inertial frame is known \cite{pet:sha:teu},
\begin{eqnarray}
\Phi_{PST} = -a t - 2 a \ln \left( 1 -{2\over{r}} \right) + b (r-1) \cos \theta,
\end{eqnarray}
where $a$ and $b$ are constants. We shall  consider for the
unperturbed potential ($A = 0$) that the  fluid  is at
rest at infinity.
Following the procedure of \cite{pet:sha:teu} we find that
 the unperturbed potential is,
\begin{equation}
\Phi_0 = - a t - 2 a \ln \left( 1- {2\over{r}} \right). \label{solution}
\end{equation}
This solution can also be found in a different way. Note that  $\Phi_{,r}$ diverges near the black
hole horizon. This is a necessary condition to avoid
 the divergence
of  the particle density  on the  horizon  \cite{pet:sha:teu}.
Therefore, near the black
hole horizon, we can neglect
$\Phi_{,p}$ when compared with $\Phi_{,r}$. It
is illustrative to consider (\ref{boxcm2})  with $A=0$
\begin{equation}
\Box_{Sch}\Phi = ([r^2-2 r] \Phi_{,r} )_{,r} +
([1-p^2] \Phi_{,p})_{,p} = 0.
\end{equation}
When near the black hole we neglect the terms
containing derivatives with respect to $p$ the equation
reduces to an equation
 for $r$ only. By considering that the temporal dependence
 is given by $-at$ this equation has
for solution (\ref{solution}). Hence, the above  imposed  condition
is valid and moreover, in
the case of fluid at rest at infinity, gives
the exact solution.

Now we shall consider the perturbation,
\begin{equation}
\Phi = \Phi_0 + A \Upsilon,
\end{equation}
where $\Upsilon$ is a function of $r$ and $p$. From  (\ref{boxcm3}) we
find,
\begin{equation}
\Box_{Sch} \Upsilon = - \Box_1 \Phi_0.
\end{equation}
Thus,
\begin{equation}
([r^2-2 r] \Upsilon_{,r} )_{,r} + ([1-p^2] \Upsilon_{,p})_{,p} = {{16 a p
(r-3)}\over{(r-2)^2}}.
\end{equation}
Again, to solve this equation we use the fact that near the black
hole we can neglect the
term containing derivatives of $p$. We find,
\begin{eqnarray}
\Upsilon =&&  C_1 + {1\over{2}}  C_2 \ln \left(1 - {2\over{r}} \right) - 4 a p \left[
{2\over{r-2}} + \ln \left(1 - {2\over{r}} \right) \right. \nonumber \\ \\ &&+ 2 \ln
\left({r\over{2}} \right) \ln(r-2) - [\ln(r-2)]^2 + 2 {\rm PolyLog}
 \left(2,1- {r\over{2}} \right) \biggl] \nonumber,
\end{eqnarray}
 with
\begin{equation}
{\rm PolyLog}(n,z) = \sum_{k=1}^{\infty} {{z^k}\over{k^n}}.
\end{equation}

To obtain the range of validity of the perturbation $A \Upsilon$,
 we  compare it with
the unperturbed function $\Phi_0$. We consider that the perturbation is valid only
when the absolute value of $A \Upsilon$ is at least a $15 \%$ of the value of
$\Phi_0$. We can take values of the constants $C_1$ and $C_2$ to enlarge the range of
the radial coordinate and still be in the required precision. For example, for $A
\approx 0.01$ and the $C_1$ and $C_2$ in the interval $(0,2)$ the range of the radial
coordinate is $r \approx (2,10)$. In Fig. 1. we show  the flux lines for the case
$A=0$ (dotted lines) and  $A \approx 0.01$ (full lines). When $A \approx 0.01$, the
streamlines that were straight lines are now curved due to the black hole
acceleration. These lines are similar to the corresponding lines for a  fluid flow in
the presence of a black hole, with the condition that the fluid velocity be constant
at infinity \cite{sha,pet:sha:teu}.

Another significant quantity is the particle density of
the fluid, $n^2=-\Phi^{,\mu}\Phi_{,\mu}$. At
 first order in $A$ we obtain
\begin{equation}
n^2 = {{a^2}\over{1 + 6Ap - {2\over{r}} - 2Apr}} -
 {{16 a^2}\over{(r-2)r^3}}
 + {{ 8 a A  C_2}\over{(r-2) r^3}} + 32 a^2
 p A {{r (r-3) + 4 \{1 +
 (r-2) \ln(r-2) \} }\over{(r-2)^2 r^3}}. \label{dencm}
\end{equation}
The  particle density (\ref{dencm}) is positive in the interval wherein the
perturbation is valid. When $A \rightarrow 0$ we obtain the correct limit for a fluid
at rest at infinity \cite{pet:sha:teu}. The fluid density contours are shown in Fig.
2. We see that near the black hole these lines are closed and more dense in the
forward direction. These contour lines are also similar to the corresponding density
lines of a fluid in the presence of a black hole, with the condition that the fluid
velocity be constant at infinity \cite{pet:sha:teu}.
 Far from the black hole these lines are open, while in \cite{pet:sha:teu}
they are closed, this clearly shows the difference between a standing black hole in a
moving fluid  and an accelerated black  hole moving in a fluid which is at rest at
infinity. The qualitative aspects of the density contours do not vary when lowering
the acceleration or varying the constant $C_2$. When $A \rightarrow 0$ we density
contours are circles centered in the black hole.

\subsection{Potential flows and  accelerated rigid spheres}

In the case of rigid spheres moving in a fluid, the particle density is no
longer divergent on the sphere surface nor $\Phi_{,r}$ is divergent. Then,
even near the surface of the sphere, we cannot neglect terms containing
derivatives of $p$ in the equation for $\Phi$. To find a numerical
 solution we need to solve the partial differential Eq. (\ref{diffeq})
  with mixed   boundary conditions, Neumann and Dirichlet. In
 this case it is more convenient to
work with the metric (\ref{cmetric}) which is
diagonal and later perform the coordinate
 transformation (\ref{trans}). In  this metric Eq.
(\ref{diffeq}) reads,
\begin{equation}
- {{\Phi_{,tt}}\over{F (q+p)^2}} + \left[{G\over{(q+p)^2}} \Phi_{,p} \right]_{,p} +
\left[{F\over{(q+p)^2}} \Phi_{,q} \right]_{,q}+ {{\Phi_{,\varphi\varphi}}\over{G
(q+p)^2}} =0.
\end{equation}
 From the condition that the fluid is stationary and
 that $\Phi$ is
independent of $\varphi$ we get,
\begin{equation}
\left[{G\over{(q+p)^2}} \Phi_{,p} \right]_{,p} + \left[{F\over{(q+p)^2}} \Phi_{,q}
\right]_{,q} =0. \label{spherecm}
\end{equation}
 The boundary conditions are: i) Zero normal
component  of the fluid velocity on the sphere surface, and
 ii) Faraway
from the sphere we  set the fluid velocity to be constant and
parallel to the acceleration. The condition i) do not describe the
typical flow of gas around a star because the surface of the star
is not hard but gaseous, except in special astrophysical
situations; but describes an idealized strong-field star.

The  code used to solve (\ref{spherecm}) is a  finite
difference multigrid method \cite{fer:per} with
second order precision. The numerical multigrid
 is evenly spaced
 in $r$, but not in $\theta$. We find that the convergence  is better
 when we  increase the number
of points in $\theta$ rather than in $r$. To solve
(\ref{spherecm}) we employ four
 domains (grids) limited by spherical shells
 of radius greater than  $2.5$ (Schwarzschild
 radius equal $2$). For the first grid we have $r\in[2.5, 3.9]$
 and we use $79300$ points; the respective numbers for
 the other three
 grids are $[3.9 ,5.3], \; 39700 $, $[5.3, 11.1],\; 19900 $,
 and $[11.1, 25.5],\; 5000$. An iterative method in which at
 every step each point of the grid
is calculated from the values of the four nearest neighbors  is
employed. The
 error is estimated with the expression
  $|\Phi^{(x_i)}_{new} - \Phi^{(x_i)}_{old}|$,
in which $\Phi^{(x_i)}_{old}$ is the old value of $\Phi$ at the
point $(x_i)$ and $\Phi^{(x_i)}_{new}$ is the new calculated value
at the same point. The program will stop when the sum of all
errors of the grid points reach some pre-establish value, say
$\sum_{x_i} |\Phi^{(x_i)}_{new} - \Phi^{(x_i)}_{old}| \leq Error$.
The code was tested with the exact solution for a rigid sphere in
a  fluid flow \cite{sha}. The numerical results on the
 surface of the sphere have less than  $1\%$ of error.
Since we are interested in qualitative aspects of the fluid behavior
rather than in precise figures this accuracy is sufficient for our
purposes. The precision in the rest of the grid is better.

In Fig. 3.  the streamlines of a fluid disturbed by a moving rigid sphere
with constant acceleration $A = 0.01$ are depicted. We see that the
streamlines concentrate in the frontal part and separate when passing
the rigid sphere. This may be due to the deformation that suffer the
killing horizons in an accelerated frame \cite{far:zim}. The qualitative
aspects of the density contours for this case are similar to the black
hole case.

\section{Potential flows in black holes and rigid spheres with dipolar
halo}

To incorporate the dipolar field into the Schwarzschild metric
 we consider the static axially symmetric  Weyl metric,
\begin{equation}
ds^2=-e^{2\nu} dt^2 + e^{2(\gamma-\nu)} \left( dz^2+d\rho^2 \right) +
e^{-2\nu} \rho^2 d\varphi^2, \label{dipole}
\end{equation}
where $\nu$ and $\gamma$ are functions of $\rho$ and $z$ only
 and
satisfy the conditions \cite{syn},
\begin{eqnarray}
&& \nu_{,\rho\rho} + {1\over{\rho}} \nu_{,\rho} +\nu_{,zz} = 0,
\label{laplace} \\ && d\gamma = \rho
\left[ (\nu_{,\rho})^2 - (\nu_{,z})^2 \right] d\rho +
 2 \rho \nu_{,\rho} \nu_{,z} dz. \label{quadrature}
\end{eqnarray}
The first equation is the usual Laplace's equation in
cylindrical coordinates, and the second (once $\nu$ is known)
 gives $\gamma$  as a quadrature.

For a multipolar expansion of  $\nu$,  the spherical coordinates
 $(r, \theta,\varphi)$ or the  prolate spherical coordinates
 $(u, v, \varphi)$ are  more adequate than the cylindrical
 ones. The relation between these coordinates are

\begin{eqnarray}
u&=&{1\over{2m}} \left[ \sqrt{\rho^2 + (z+m)^2} +\sqrt{\rho^2
+(z-m)^2} \right], \nonumber \\
&=&{r\over{m}}-1, \; u \geq 1, \nonumber \\
v&= &{1\over{2m}} \left[ \sqrt{\rho^2 + (z+m)^2} -\sqrt{\rho^2
+(z-m)^2} \right], \\
&=&\cos{\theta}, \; -1 \leq v \leq 1, \nonumber \\
\varphi& = &\varphi. \nonumber
\end{eqnarray}

 In terms of $\rho$ and $z$,
\begin{eqnarray}
\rho & = & m \sqrt{(u^2-1)(1-v^2)} \nonumber \\
&= &\sqrt{r(r-2m)} \sin{\theta}, \; r \geq 2m, \nonumber\\
z& = & m uv  \label{trans2} \\
&=&(r-m) \cos{\theta}, \nonumber \\
\varphi &= &\varphi. \nonumber
\end{eqnarray}

\noindent As before, we will set $m=1$.

Using the transformations (\ref{trans2}), Eqs. (\ref{laplace}) and
(\ref{quadrature}) can
be written in terms of $u$ and
$v$ as
\begin{eqnarray}
&& \left[ (u^2-1) \nu_{,u} \right]_{,u} + \left[(1-v^2) \nu_{,v} \right]_{,v} = 0,
\\ && \gamma_{,u} = {{1-v^2}\over{u^2-v^2}} \left[ u(u^2-1) (\nu_{,u})^2 -u(1-v^2)
(\nu_{,v})^2 - 2v(u^2-1) \nu_{,u} \nu_{,v} \right], \nonumber
\\ && \gamma_{,v} = {{u^2-1}\over{u^2-v^2}} \left[ v(u^2-1) (\nu_{,u})^2 -v(1-v^2)
(\nu_{,v})^2 + 2u(1-v^2) \nu_{,u} \nu_{,v} \right] \nonumber.
\end{eqnarray}
The authors in \cite{vie:let} solved this equations using a
external multipolar expansion up to octopoles using a Legendre expansion
with their corresponding terms increasing with the distance in the
intermediate vacuum between the core and the shell. In this work we
 are only interested in the dipolar case. Hence, the  functions $\nu$ and
$\gamma$ reduce to

\begin{eqnarray}
2\nu &=& 2 \nu_0 +\kappa \ln \left( {{u-1}\over{u+1}}
\right)+ 2{\cal D} uv , \label{funcnu} \\
2\gamma &=& 2 \gamma_0 + \kappa^2 \ln \left(
 {{u^2-1}\over{u^2-v^2}} \right) + 4
\kappa {\cal D}v - {\cal D}^2 \left[ u^2(1-v^2) + v^2 \right],
\label{funcgamma}
\end{eqnarray}
where $\cal D$ is the value of the dipolar field produced
 by an axially
symmetric halo (or shell) of matter, and $\kappa$ is a constant. The Schwarzschild
solution is recovered with $\kappa=1$ and ${\cal D} =0$. The constants $\gamma_0$ and
$\nu_0$ can be used to rule out conical singularities and to ensure analyticity of the
metric at the horizon \cite{vie:let};  here we have made them zero. From the metric
(\ref{dipole}) in coordinates $u$ and $v$, and (\ref{funcnu}) and (\ref{funcgamma}),
we find
\begin{eqnarray}
ds^2 &=& -\left( {{u-1}\over{u+1}} \right)
e^{2{\cal D}uv} dt^2 + (u+1)^2 e^{2{\cal
D}v(2-u) - {\cal D}^2 [u^2(1-v^2) +v^2]}
 \left[ {{du^2}\over{u^2-1}} \right. \nonumber
\\ && + \left. {{dv^2}\over{1-v^2}} \right]
+ (u+1)^2(1-v^2) e^{-2{\cal D}uv}
d\varphi^2. \label{dipoleuv}
\end{eqnarray}
 This metric represents a
  monopolar core ($\kappa =1$) in the
presence of a external dipolar field $({\cal D})$ that
 is associated to a distant
shell or halo of matter.

\subsection{Potential flow of a  rigid
 sphere in a external dipolar field}

For a  stationary fluid  that does not depend
on $\varphi$ the scalar equation
(\ref{diffeq}) reduces to
\begin{equation}
\left[ (u^2-1) \Phi_{,u} \right]_{,u} +
\left[(1-v^2) \Phi_{,v} \right]_{,v} = 0. \label{diffuv}
\end{equation}
This is exactly the same equation that we solved in Sec. III.A, but in a
different system of coordinates.  Note that all the Weyl solutions have
the same differential equation (\ref{diffuv}) for $\Phi$.
The  presence of the dipolar field
will be taken into account in
the boundary conditions. On the surface of the rigid sphere
the boundary condition is the usual one
 for the Euler's equation, i.e., that
the normal component of the  fluid velocity  vanish on the surface.
For large values of $u$ the boundary
condition deserves more attention.
  Since, the time dependence
of $\Phi$ is $-at$, we have for the density,
\begin{equation}
n^2= a^2
\left({{u+1}\over{u-1}} \right) e^{-2{\cal
D}uv} - {{e^{2{\cal D} v(u-2) + {\cal D}
 ^2[u^2(1-v^2)+v^2]}}\over{(u+1)^2}} \left[
(\Phi_{,u})^2 (u^2-1) + (\Phi_{,v})^2 (1-v^2) \right]. \label{dendipole}
\end{equation}
Due to the presence of the dipolar field ${\cal D}$,
at some value of $u$ greater than
a certain $u=u_c$, the first term in (\ref{dendipole}) will always be
 smaller than the
others two. Then we will have $n^2 < 0$.
This is not allowed; we say that the fluid
is no longer stationary, i.e., the assumed time dependence
is not right for $u>u_c$. If ${\cal D}$ is large
 enough, say ${\cal D} \approx 1$,
the fluid will never be stationary. The first term also
depends on $a^2$. By varying the value of $a$ we can enlarge
or decrease
 the domain where the fluid is stationary, also for ${\cal D}
\rightarrow \infty$ we can
set $a$ large enough to keep the fluid stationary.
 For example, for $a=1.25$, we must have
${\cal D} \approx 0.001$ to obtain a relative large
domain wherein the fluid
remains stationary. Test particles moving in the metric
(\ref{dipoleuv}) can have chaotic behaviour \cite{vie:let}.
We believe that the lost of the stationary character
of the fluid for $u>u_c$ may be   another manifestation of
the same type of  instability. This point is under active
consideration by the authors.

The  outer boundary condition
is still missing. With all
these fluid instabilities it is not clear which
  is the right boundary condition.
The previous condition, that
the fluid velocity  to be constant far from the sphere,
 is not a valid condition in this case because
the fluid is accelerating and is in this region where the
instabilities appear. To find some characteristics
values of $\Phi$ to be used
as an outer
boundary condition we will integrate $\Phi_{,u}$ along a line of constant
$v$  from the surface of the sphere (black hole) to
 the stationary limit of the fluid. From Eq. (\ref{velocity}) we get
\begin{equation}
\Phi_{,u} \propto {{g_{uu} U^u}\over{g_{tt} U^t}}.
\end{equation}
The proportionality constant is of no importance,
because $U_\mu$ in (\ref{velocity}) is invariant under a rescaling of
$\Phi$. To find the $t$ component of the four-velocity, $U^t$,
we will assume -- without further justification -- that
the a fluid particle follows  the usual
geodesic equation of motion for a test particle, i.e., the
Euler-Lagrange equation,
\begin{equation}
{d\over{ds}} \left( {{\partial L}\over{\partial \dot{x}^{\mu}}} \right) -
{{\partial L}\over{\partial x^{\mu}}} = 0,
\end{equation}
where  $\dot{x}^{\mu}$ represents total derivative of
$x^{\mu}$ with respect to the parameter $s$ and
 $L(\dot{x}^\lambda,x^\lambda) ={1\over{2}} g_{\mu\nu} \dot{x}^\mu
\dot{x}^\nu$. Since the coordinate $t$ is cyclic,
 we find,
\begin{equation}
\dot{t}=U^t=- k \left( {{u+1}\over{u-1}} \right) e^{-2{\cal D}uv},
\end{equation}
where the constant $k$ is the energy function of
the fluid particle.
 The component $U^u$ along the lines of constant $v$ can be computed
 from the metric (\ref{dipoleuv}),
\begin{equation}
U^u = \dot{u}= \sqrt{  \left( {{u-1}\over{u+1}} \right) e^{2 {\cal D} uv}
+k^2} \; e^{ -2 {\cal D}v + {{\cal D}^2\over{2}} [u^2 (1-v^2) +v^2]}
\end{equation}
 Therefore, $\Phi$ along the lines of  constant $v$ can be written as
\begin{equation}
\Phi = \int \left( {{u+1}\over{u-1}} \right)
e^{2 {\cal D}v(1-u)-{{{\cal
D}^2}\over{2}}[u^2(1-v^2)+v^2]} \sqrt{ \left(
{{u-1}\over{u+1}} \right) e^{2 {\cal
D}uv} + k^2} \; du, \label{integral}
\end{equation}
where  a multiplicative constant was set equal to one. We numerically
integrate (\ref{integral}) to
obtain the outer boundary condition at a certain value $u$ near
the limit of stability. We use the code  described in Sec. III.B
to solve (\ref{diffuv}). For different values of the constant $k$
the qualitative features of the streamlines and density contours
do not change. In Fig. 4  the streamlines of a fluid in the presence of a
rigid sphere of radius $2.5$, Schwarzschild radius equal 2, and a
external dipolar moment field with ${\cal D}=0.001$ are shown.  We also
set $k=1$ in this case. The values of $\Phi$ used  for boundary condition
and  all the ones computed satisfy the condition $n^2>0$. We see little
difference in the flows with ${\cal D}=0.001$ and ${\cal D}=0$. In Fig. 5
the density contours are plotted for the same values of the parameters
used in Fig 4, sphere of radius $2.5$ and ${\cal D}=0.001$. We see the
same qualitative features discussed in Sec. II.A in the context of the
C-metric.  In this case the fluid is accelerating something that we should
expect since an external field (in this case the external dipolar
field) exert a force on the fluid.

\subsection{Potential flows for black holes with dipolar halo}

Like in the precedent  case  we solve Eq. (\ref{diffuv}),
but to avoid the singularity at the
black hole horizon, we  use the
tortoise radial coordinate \cite{reg:whe},
\begin{eqnarray}
r^*&=&r+2 \ln \left( {r\over{2}} -1 \right) \\
&= &(u+1)+2 \ln \left( {{u-1}\over{2}}
\right),
\end{eqnarray}
Now, Eq. (\ref{diffuv}) reads,
\begin{equation}
\left( {{u+1}\over{u-1}} \right) \left[ (u+1)^2 \Phi_{,r^*} \right]_{,r^*} +
\left[(1-v^2) \Phi_{,v}  \right]_{,v} = 0,
\end{equation}
and the particle density takes the form,
\begin{equation}
n^2=a^2 \left( {{u+1}\over{u-1}} \right)
e^{-2{\cal D} uv} - e^{2{\cal D}v(u-2) +
{\cal D}^2 [ u^2 (1-v^2) +v^2]} \left[
 {{u+1}\over{u-1}} (\Phi_{,r^*})^2 +
{{1-v^2}\over{(u+1)^2}} (\Phi_{,v})^2 \right].
\end{equation}

From a computational  view point  the difference between a rigid
sphere
 and a black hole laid in the inner
boundary condition. For a rigid sphere the condition
 is  zero normal velocity  on its surface,
and for a black hole is finite particle
 density  on the
 horizon. This last condition
   requires that $\Phi_{,r^*}$ be limited in such
a way that cancels the singular term
 in $\Phi_{,v}$.
In the general time dependent
case this condition reads \cite{abr:sha,fon:mar:iba:mir},

\begin{equation}
\Phi_{,r^*} = \left(1- {2\over{r}} \right) \Phi_{,r}
= \Phi_{,t} + a_1(t)(r-2) +
a_2(t)(r-2)^2 + \cdots, \label{expan}
\end{equation}
where $a_1, a_2, ...$ are functions of time coordinate.
Hence, the following equation is valid near $r=2$
\begin{equation}
{\partial\over{\partial r^*}} \left[ {{\Phi_{,r^*}
-\Phi_{,t}}\over{r-2}} \right] =0. \label{condtort}
\end{equation}

With the
 assumption that
$\Phi_{,t}=-a$ this condition is also valid in our case and
 it will be taken as the
 inner boundary
condition. In (\ref{expan}) we cannot set  $\Phi_{,r^*} =
\Phi_{,t}$ because we can have $n^2\leq 0$. The outer condition
for this problem is found integrating Eq. (\ref{integral}) as
before. The numerical code employed is the same of Sec. IV.A with
the implementation of the inner boundary
 condition (\ref{condtort}) and the change of $r$ or $u$ to the tortoise
coordinate $r^*$. The program was tested
with the exact solution for a black hole in a moving fluid
\cite{pet:sha:teu}.
 The computed values of $\Phi$  on
the black hole horizon have less than $1 \%$
 of error when compared with the exact solution. Again,
 outside the horizon the precision is
better.

In Fig. 6 the streamlines of a perfect fluid in the presence of a black
hole in an external dipolar moment field  with ${\cal D}=0.001$  are
plotted. Like the rigid sphere case the streamlines show little difference
with respect to the case ${\cal D}=0$. In Fig. 7. we present the density
 contours for the same values of the parameters used in the previous
figure.  Once more we see the same pattern encountered in the C-metric case. This
shows that the effect of acceleration appears always in the density contours rather
than in the fluid streamlines. In the last four figures  we see that the streamlines
are practically the same compared to the respective case with ${\cal D}=0$, but the
density contours are quite different. The density contours for the different obstacles
present remarkable similar features. We think that this pattern may be taken as test
for the presence of acceleration in this kind of fluids.

\section{Conclusions}

We can summarized the results of this work as follows: in the C-metric and dipolar
field cases we found that, for a stationary fluid, the difference between the
accelerate and non accelerate cases lies on the form of the density contours and not,
as one may think, in the shape of the streamlines. We assumed in both cases that the
acceleration is near zero. In the dipolar field case the potential flow becomes very
unstable and the presence of chaos may appear \cite{vie:let}.

It is important to note that the method presented hear is valid for a stiff equation
of state and the incorporation of a new barotropic equation of state for the fluid is
not easily applied. In that case we need a different approach for solving the
difference equation (\ref{eqphi}) because it is nonlinear and it is necessary to
impose initial conditions to the enthalpy, this may help or not in the convergence of
the method. A negative value of the enthalpy could indicate a nonstationary regime for
the fluid , see for instance \cite{abr:sha}. In the different scenarios of this work
the search for suitable boundary conditions to be treated, with $A \approx 0$, have
been a constant problem, we think that this difficulties would increase with a new
barotropic equation due to the condition for the enthalpy. Future applications of this
work could be the study of fluids with another equation of state to model a more
realistic situation, the study of chaos in the presence of instabilities, and the
study of fluids in non stationary states. All this applications are currently under
investigation by the authors.

\section*{Acknowledgements}

We want to thank FAPESP and CNPq for financial support, M.U. also
thanks Sebastian Ujevic for discussions through all the preparation
of this work.


\newpage
\begin{figure}
\centering \vbox to 9.0cm {\vss\hbox to 9cm{\hss\ {\includegraphics{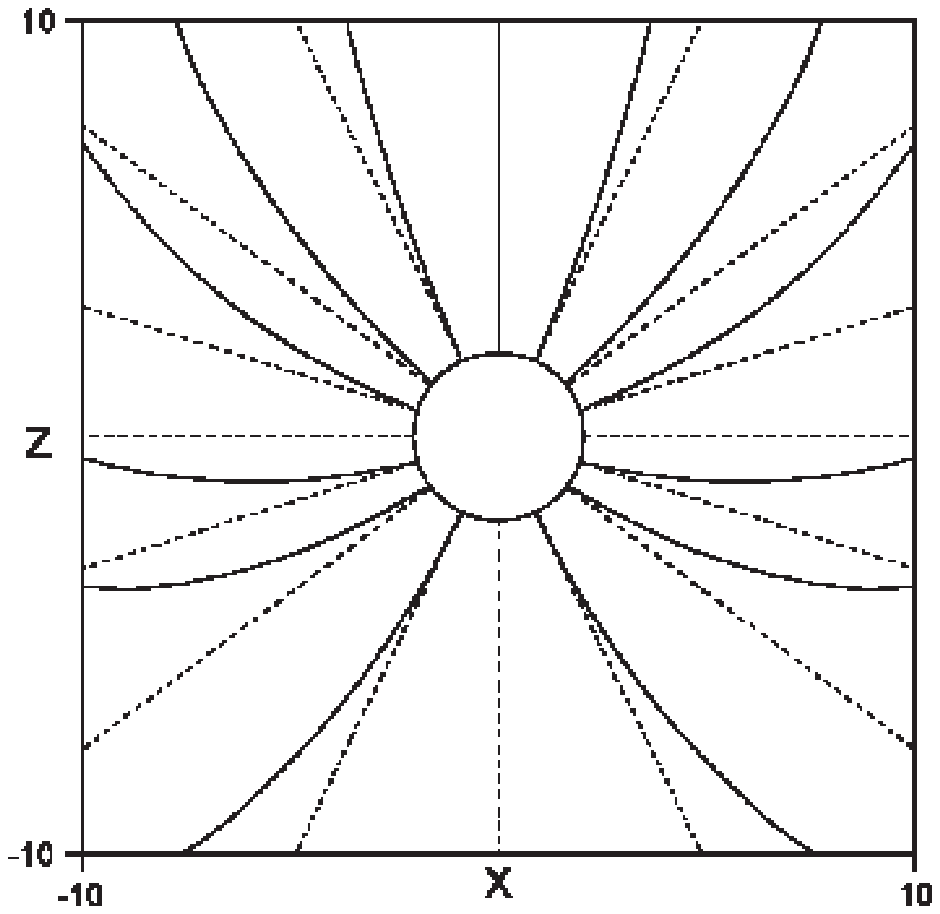}}\hss}}
\caption{Analytical results for the streamlines of a fluid in the presence  of an
accelerating  black hole in the direction of the positive $z$-axis. The solid lines
represent the case with $A \neq 0$ and the dotted lines $A=0$. The black hole has
radius $r \approx 2$ (Schwarzschild radius equal 2), this approximation is due to the
deformation of the Schwarzschild horizon. The axes are defined as $X=r \sin \theta$
and $Z=r \cos \theta$, with $r=1/[A(p+q)]$ and $\sin^2 \theta =G(p)$.}
\end{figure}

\begin{figure}
\vbox to 9.0cm {\vss\hbox to 9cm{\hss\ {\includegraphics{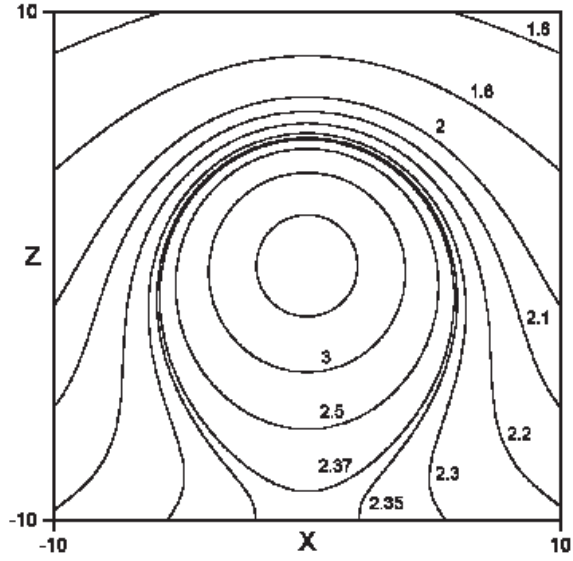}}\hss}}
\caption{Analytical results of the fluid  density contours of an accelerating black
hole. The black hole radius and the meaning of the axes are defined as in Fig. 1.}
\end{figure}

\begin{figure}
  \vbox to 9.0cm {\vss\hbox to 9cm{\hss\ {\includegraphics{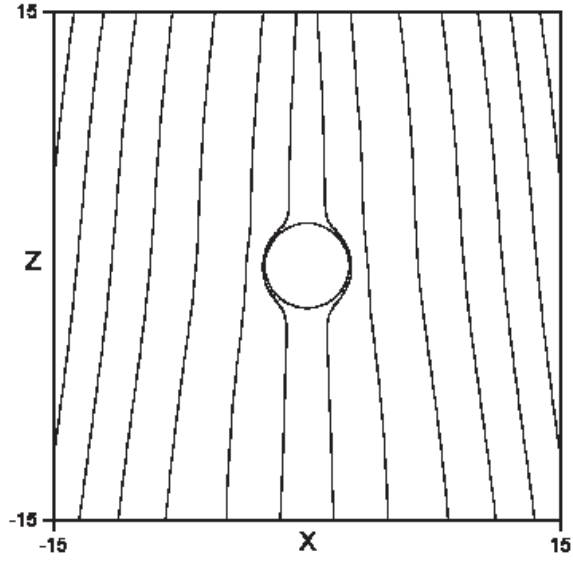}}\hss}}
\caption{Numerical results for the fluid streamlines in the presence of a rigid sphere
accelerating in the direction of the positive $z$-axis. The sphere radius is $r
\approx 2.5$ (Schwarzschild radius equal 2), this approximation is due to the
deformation of the Schwarzschild horizon. The axes are defined as in Fig. 1}
\end{figure}

\begin{figure}
  \vbox to 9.0cm {\vss\hbox to 9cm{\hss\ {\includegraphics{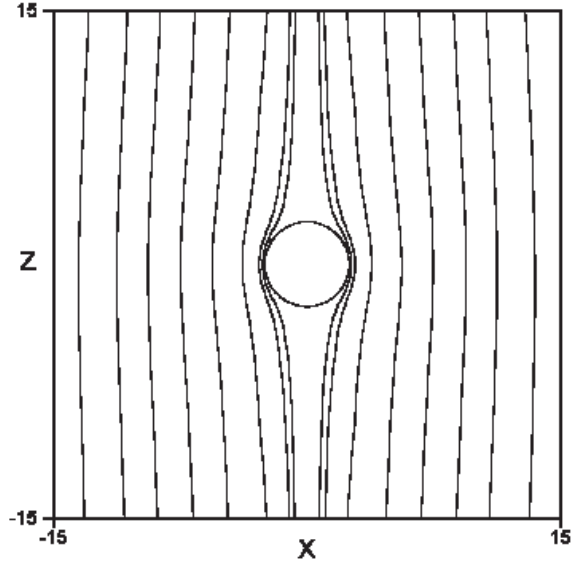}}\hss}}
\caption{Numerical results for the fluid streamlines when an external dipolar field
with ${\cal D}=0.001$ is present and a rigid sphere of radius $ r = 2.5$
(Schwarzschild radius equal 2) is placed as an obstacle. The axes are defined as $X=r
\sin \theta$ and $Z=r \cos \theta$, with $r=u+1$ and $\cos \theta = v$. }
\end{figure}

\begin{figure}
  \vbox to 9.0cm {\vss\hbox to 9cm{\hss\ {\includegraphics{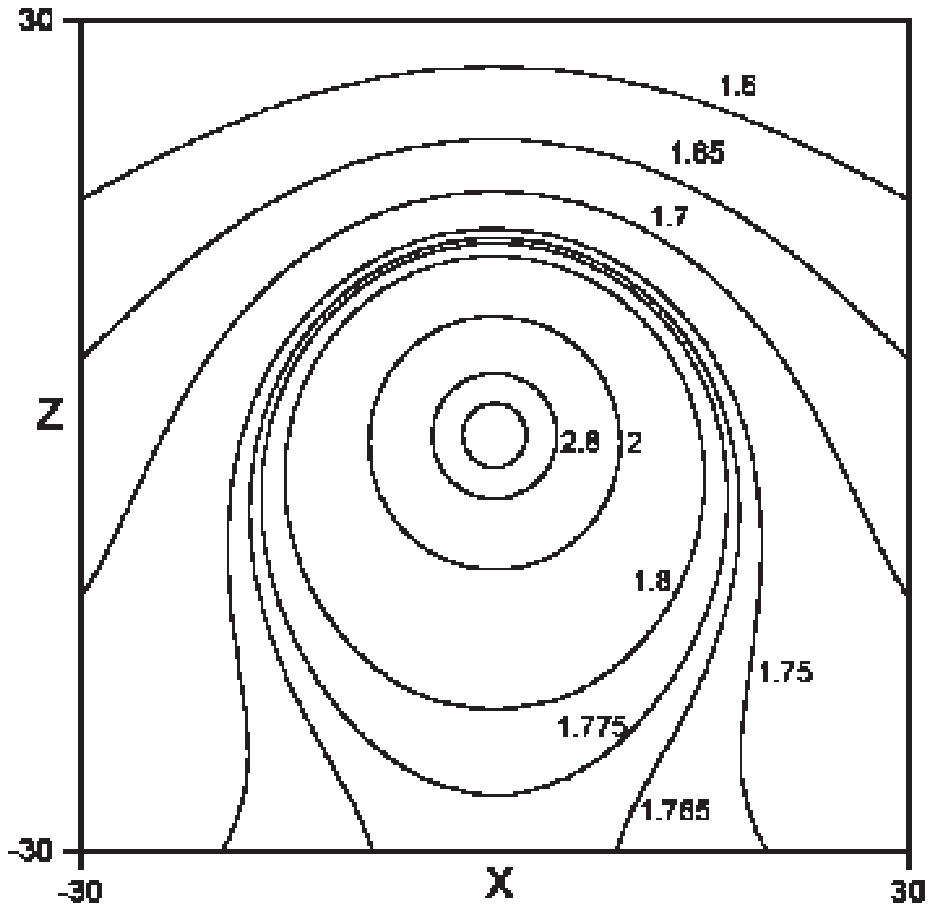}}\hss}}
\caption{Numerical results for the fluid density contours when an external dipolar
field with ${\cal D}=0.001$ is present  and a rigid sphere of radius $r = 2.5$
(Schwarzschild radius equal 2) is placed as an obstacle. The axes are defined as in
Fig. 4.}
\end{figure}

\begin{figure}
  \vbox to 9.0cm {\vss\hbox to 9cm{\hss\ {\includegraphics{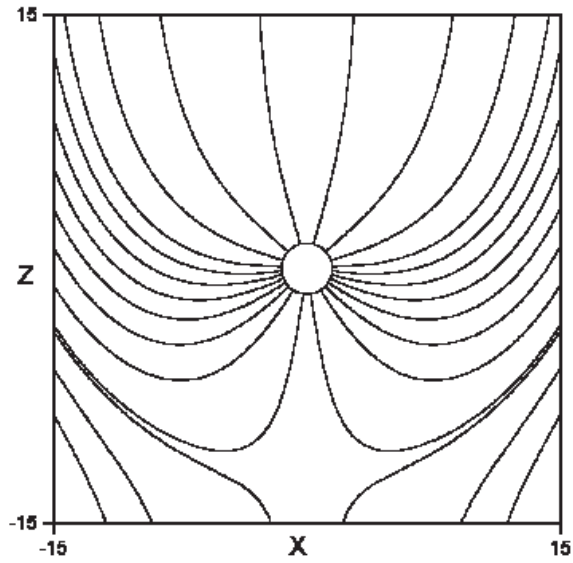}}\hss}}
\caption{Numerical results for the streamlines when an external dipolar field of value
${\cal D}=0.001$ is present and a black hole is placed as an obstacle. The black hole
has radius $r=2$ (Schwarzschild radius equal 2). The meaning of axes are the same of
Fig. 4.}
\end{figure}

\begin{figure}
  \vbox to 9.0cm {\vss\hbox to 9cm{\hss\ {\includegraphics{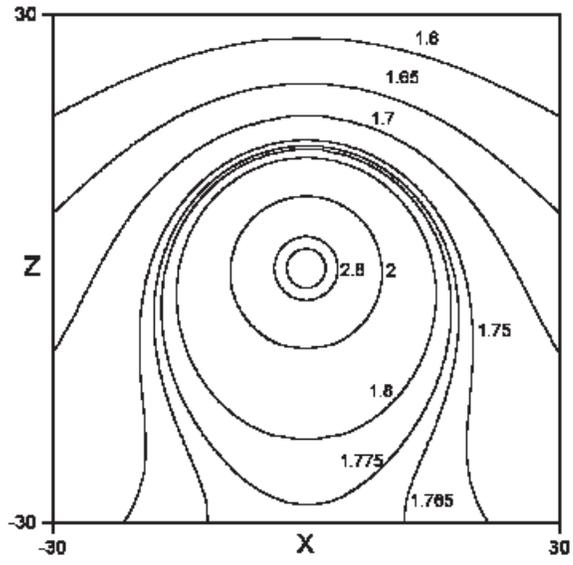}}\hss}}
\caption{Numerical results for the fluid density contours when an external dipolar
field of value ${\cal D}=0.001$ is present and a black hole is placed as an obstacle.
The black hole has radius $r=2$ (Schwarzschild radius equal 2). The meaning of the
axes are the same of Fig. 4.}
\end{figure}

\end{document}